\documentclass[12pt]{article}

 \usepackage{graphicx}
 \usepackage{psfrag}
 \usepackage{flafter}

 \newcommand {\nc}{\newcommand}
 \nc{\be}{\begin{equation}}
 \nc{\ee}{\end{equation}}
 \nc{\bea}{\begin{eqnarray}}
 \nc{\eea}{\end{eqnarray}}
 \def\complex{{\mathcal C}}
 
 \def\integer{{\mathcal Z}}

 \def\unit{{\mathbf 1}}
 \def\order{{\mathcal O}}
 
 \def\H{{\mathcal H}}
 
 \def\pr{\prime}
 \def\tld{\tilde}
 \def\bra{\langle}
 \def\ket{\rangle}
 \def\LC{\delta\sigma}
 \def\vec{\underline}
 \def\lp{\hat{\partial}}

 \newtheorem{theorem}{Theorem}
 \newtheorem{lemma}{Lemma}

 \title{Quiver Matrix Mechanics for IIB String Theory (II):
 Generic Dual Tori, Fractional Matrix Membrane and
 $SL(2,\integer)$ Duality}
 \author{
  Jian Dai\thanks{jdai@physics.utah.edu},
  Yong-Shi Wu\thanks{wu@physics.utah.edu}\\
  Department of Physics, University of Utah\\
  Salt Lake City, Utah 84112}

 \date{\today}

\begin{document}
  \maketitle


  \begin{abstract}

   With the deconstruction technique, the geometric information
   of a torus can be encoded in a sequence of orbifolds. By
   studying the Matrix Theory on these orbifolds as quiver
   mechanics, we present a formulation that (de)constructs the
   torus of {\em generic shape} on which Matrix Theory is
   ``compactified''. The continuum limit of the quiver mechanics
   gives rise to a $(1+2)$-dimensional SYM. A hidden (fourth)
   dimension, that was introduced before in the Matrix Theory
   literature to argue for the electric-magnetic duality, can be
   easily identified in our formalism. We construct membrane
   wrapping states rigorously in terms of Dunford calculus in
   the context of matrix regularization. Unwanted degeneracy
   in the spectrum of the wrapping states is eliminated by using
   $SL(2,\integer)$ symmetry and the relations to the FD-string
   bound states. The dual IIB circle emerges in the continuum
   limit, constituting a critical evidence for IIB/M duality.
  \end{abstract}


  \section{Introduction}
  \label{section_int}


   The duality between type IIB string theory and M-theory,
   as an indispensable component of the string/M-theory web,
   {\it \`{a} la} Schwarz and Aspinwall \cite{SCH,SCH1,AS},
   has two characteristic features: First there is a
   generalized T-duality between IIB theory on a circle
   and M-theory one a two-torus. According to this duality, the
   winding modes of the fundamental/D-string bound states (FD-string) in the IIB spectrum are
   dualized to Kaluza-Klein (KK) modes in M-theory. Secondly
   the non-perturbative $SL(2,\integer)$ symmetry in IIB
   theory is geometrized to be the modular group of the
   two-torus in M-theory. The physical ideas behind this
   duality are elegant and beautiful, but how to formulate
   them in an explicit formalism and in a constructive way
   remains a challenge.

   To our knowledge, up to now the only way available to
   formulate M-theory microscopically is the BFSS Matrix
   Theory \cite{BFSS}. (For other attempts to formulate
   the notion of the so-dubbed ``protean degrees of freedom"
   of M-theory, see the review \cite{BS_R}.) In this
   framework the statement for the non-perturbative IIB/M
   duality is that Matrix Theory compactified on a two-torus,
   with the size of the torus shrinking to zero, is dual to
   IIB string theory in a flat background \cite{S,SS,FHRS}.
   However, it is highly non-trivial to see how this can
   come about. Matrix theory has nine transverse dimensions;
   when two of them are compactified and shrink to vanishing
   size, only seven dimensions survive. One needs to have
   the eighth transverse dimension in IIB string theory
   emerging in this limit. This emergent dimension should
   have two crucial properties in conformity with the
   IIB/M duality. First before decompactification the KK
   modes along it have to be associated with the membrane
   wrapping states on the torus in Matrix Theory. Secondly
   after decompactification it should be on the same
   footing as other transverse dimensions, sharing an
   eight dimensional rotational invariance.

   Since the compactified Matrix
   Theory is formulated as a $(1+2)$-dimensional
   Super Yang-Mills theory (SYM$_{1+2}$) on the dual torus,
   IIB/M(atrix) duality is
   addressed in the language of SYM in refs. \cite{S,SS,FHRS}.
   On the one hand, the wrapping membranes are argued to
   correspond to configurations of the Yang-Mills fields,
   with nonvanishing (abelian) magnetic flux, which gives
   the wrapping number. On the other hand, the rotational
   symmetry between the decompactified emergent dimension
   and other flat transverse ones in the IIB target space,
   is argued to be related to the (conjectured)
   electric-magnetic (EM) duality of the $(1+3)$-dimensional
   SYM, resulting from Matrix Theory compactified on a
   three-torus. Though these intuitive SYM arguments are
   compelling, an explicit construction of the wrapping
   membrane, as well as the definition of its wrapping
   number, is in demand in either the compactified Matrix
   Theory or in the dual $(1+2)$-dimensional SYM \cite{SCH2}.
   Certainly one would prefer a more direct approach without
   the detour into the EM-duality in the three-dimensional
   world-volume of SYM. Moreover, IIB/M(atrix) duality has
   been addressed only for the rectangular tori. No serious
   attempts have been made to formulate the notion of ``dual
   torus'' of {\sl generic shape}.


   To fill the gap, in our previous work in this series
   \cite{DW1}, we tried to generalize the definition
   of the wrapping number for a continuous map between
   two tori to matrix states wrapping on the compactified
   torus. We first adopted the deconstruction techniques
   \cite{ACG} to approximate the compactified torus by
   a sequence of orbifolds, that encode the geometric
   information of a {\it rectangular} torus. This resulted
   in a quiver matrix quantum mechanics, whose continuum
   limit gives rise to SYM$_{1+2}$. Then
   the wrapping matrix states were constructed explicitly
   in the quiver matrix mechanics framework in terms of
   fractional powers of the 't Hooft clock and shift
   matrices. And it has been checked that this construction
   possesses all properties required by the IIB/M duality.
   The present paper is sequential to the previous one,
   to study the case in which the compactified torus is
   of {\it generic} shape. Our motivation is to better
   understand how the {\em generic} geometry of the {\it
   dual torus} is encoded in the discrete setting of
   deconstruction (based on the orbifolding approach),
   and how a continuous geometry is restored on the SYM
   world-volume in the continuum limit. We shall refine
   the formalism for the fractional powers of matrices
   by employing a functional calculus of Dunford, to
   construct matrix membrane states with a
   well-defined topological wrapping number. Then, the
   generalized (torus-circle) T-duality is verified for
   generic moduli of the torus on a rigorous ground.
   As a bonus, we shall gain some insights into the
   $SL(2,\integer)$ duality by identifying the
   FD-string in M-theory and eliminating the unwanted
   degeneracy in the membrane wrapping states in a
   satisfying fashion.


   The outline of this paper is as follows. In
   Sec.~\ref{section_orb}, we describe in detail the
   Matrix Theory on $\complex^3/\integer_N^2$ orbifold,
   which is our starting point, in the form of quiver
   matrix mechanics. In Sec.~\ref{section_dec}, we
   (de)construct in the large $N$ limit the toroidal
   world-volume geometry with generic moduli, with a
   $(1+2)$-dimensional SYM defined on it, which recovers
   Matrix Theory compactification on a dual torus.
   Moreover, we are able to identify the projection of
   the $(1+3)$-dimensional SYM involved in previous
   discussions in the literature on IIB/M(atrix) duality.
   Since the geometric information of the target torus
   is (re)constructed from the orbifold data, this
   results in a precise formulation of the duality between
   the target torus and the world-volume torus of SYM. In
   Sec.~\ref{section_mem}, we suggest a matrix construction
   of the membrane wrapping on the compactified torus, by
   using {\em Dunford calculus}. The IIB spectroscopy
   including the bound FD-string is analyzed, with the
   unwanted degeneracy of the wrapping states eliminated
   as expected. Moreover the $SL(2,\integer)$ symmetry
   among the wrapping states is presented in the general
   setting. In Sec.~\ref{section_dis}, we dwell upon remarks,
   problems and a few perspectives. The relationship of
   this work to various approaches in the literature is
   discussed too.

  \section{Quiver Matrix Mechanics from Matrix Theory on
      $\complex^3/\integer_N^2$}
  \label{section_orb}

   In the IIB/M(atrix) duality, the dynamics on the M-Theory
   side is described by the Matrix Theory compactified on a
   two-torus. We (de)construct the two-torus in terms of the
   orbifold $\complex^3/\integer_N^2$, which is defined in
   the following way. Parameterize $\complex^3$ by three complex
   numbers $z^a$, $a=1,2,3$; the actions of the discrete group
   $\integer_N^2$ on $\complex^3$ introduces two classes of
   equivalence relations
   \bea
   \label{ORBIFOLD}
   \nonumber
    \mbox{I:}
    &&
    z^1\sim \omega_N^\ast z^1,
    z^2\sim \omega_N z^2,
    z^3\sim z^3;\\
    \mbox{II:}
    &&
    z^1\sim \omega_N^\ast z^1,
    z^2\sim z^2,
    z^3\sim \omega_N z^3
   \label{OC}
   \eea
   where $\omega_N = e^{i2\pi/N}$.

   The action of $U(K)$ Matrix Theory on this orbifold
   reads
   \bea
    \nonumber
    S &=& \int dt Tr \{
     {1\over 2R_{11}}[D_t, Y^i]^2 + \frac{R_{11}}{4}[Y^i,Y^j]^2 \\
    \nonumber
     && + {1\over R_{11}}[D_t, Z^a][D_t, Z^{a\dag}]
     + \frac{R_{11}}{2} ([Z^a,Z^{a^\pr \dag}][Z^{a\dag},Z^{a^\pr}]
     + [Z^a,Z^{a^\pr}][Z^{a\dag},Z^{a^\pr \dag}]) \\
    \nonumber
     &&  + R_{11}[Y^i,Z^a][Y^i,Z^{a\dag}] \\
     && - {i\over 2}\Lambda^{\dag}[D_t,\Lambda]
     + \frac{R_{11}}{2} \Lambda^{\dag} \gamma_i [Y^i,\Lambda] +
     \frac{R_{11}}{\sqrt{2}}\Lambda^{\dag} (\tld{\gamma}_a [Z^a,\Lambda]
     + \tld{\gamma}_a^{\dag} [Z^{a\dag},\Lambda] )
     \}.
   \label{OA}
   \eea
   In Eq.~(\ref{OA}), the eleven-dimensional planck length is taken
   to be unity, $R_{11}$ is the radius of the compactified light-cone
   in the infinite momentum frame (IMF), $t$ the world-line time;
   $D_t = d/dt + i[A_0,.]$ with $A_0$ the $U(K)$ gauge connection in
   the temporal direction; both indices $i$ and $a$ run from $1$ to
   $3$ and a representation of the gamma matrices is given by
   \bea
    \nonumber
    \gamma_1 = -\tau_2 \otimes \tau_3 \otimes \tau_3 \otimes \tau_3&,&
    \gamma_2 = -\tau_1 \otimes \unit \otimes \tau_3 \otimes \tau_3, \\
    \nonumber
    \gamma_3 = -\tau_3 \otimes \unit \otimes \tau_3 \otimes \tau_3&,&
    \tld{\gamma}_1 = -\epsilon \otimes \tau_- \otimes \unit \otimes \tau_3,\\
    \label{GMM}
    \tld{\gamma}_2 = i\unit \otimes \tau_3 \otimes \tau_- \otimes \tau_3&,&
    \tld{\gamma}_3 = -i \unit \otimes \unit \otimes \unit \otimes \tau_-,
   \eea
   in which $\tau_{1,2,3}$ are conventional Pauli matrices and
   $\tau_- = (\tau_1 - i \tau_2)/2$.

\begin{figure}[hbtp]
    \centering
    \psfrag{l1}[][]{$\Lambda^{100}$}
    \psfrag{l2}[][]{$\Lambda^{010}$}
    \psfrag{l3}[][]{$\Lambda^{001}$}
    \psfrag{l0}[][]{$\Lambda^{000}$}
    \psfrag{Z1}[][]{$Z^1$}
    \psfrag{Z2}[][]{$Z^2$}
    \psfrag{Z3}[][]{$Z^3$}
    \psfrag{Z0}[][]{$A_0$, $Y^{1,2,3}$}
    \includegraphics[width=0.60\textwidth]{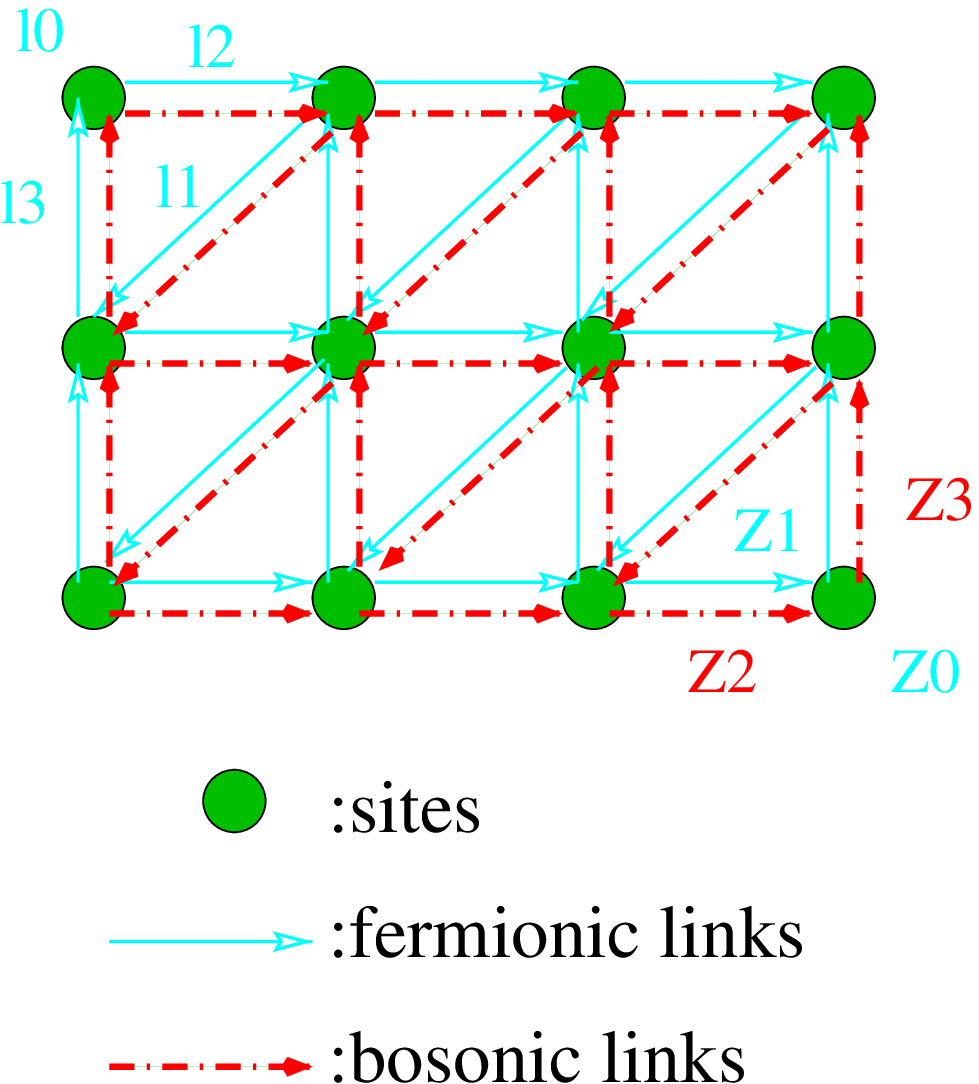}
    \caption{Quiver diagram for the Matrix Theory on $\complex^3/\integer_N^2$
    }
    \label{QDG}
   \end{figure}

   $Y^i$ and $Z^a$ are the coordinates of $K$ D-particles;
   $\Lambda$ their fermionic partner, which is an $SO(9)$
   Majorana spinor with $16$ real components.
   According to the tensorial decomposition in Eq.~(\ref{GMM}),
   the components of the fermionic coordinate are denoted
   as $\Lambda^{s_0s_1s_2s_3}$ for $s_c=0,1$, $c=0,1,2,3$,
   in which $\Lambda$ is real for $s_0$ and
   $\Lambda^{0\dag} = \Lambda^1$ for the other $s_c$.
   Because of the stringy nature of D-branes and the
   orbifold actions, all of these coordinates are lifted
   to be $KN^2\times KN^2$-matrices. Regarding the
   $N^2\times N^2$ indices from orbifolding, their
   transformation properties under the gauge symmetry
   of each variable can be directly read off from the
   quiver diagram in Fig.~\ref{QDG}, in which only six
   unit cells are presented (prolongable in two directions
   to give $N\times N$ unit cells), and  $\Lambda$ is
   labelled only by $s_{1,2,3}$; the orbifold conditions
   in Eqs.~(\ref{OC}) are automatically incorporated
   in these transformation rules. Note that all of the
   variables can be interpreted to reside in the orbifolding
   group $\integer_N^2$; therefore, the terms ``site" and
   ``link" in Fig.~\ref{QDG} are understood in the
   circumstance of the discrete group $\integer_N^2$,
   which here can be viewed as an approximated (or
   discretized) world-volume. So, sites are labelled by
   the elements in $\integer_N^2$ (pairs of integers $(m,n)$,
   with $m$, $n$ modulo $N$). In the jargons of quiver theory,
   here from the target space point of view, site variables
   are adjoint matters while link variables are bi-fundamental
   matters.

   The orbifolded Matrix Theory in Eq.~(\ref{OA}) provides the
   basic machinery to generate, in the large $N$ limit, both
   the geometry of the SYM world-volume and that of the torus
   in the target space. The next two sections are devoted to
   show how the geometries emerge in the continuum limit.

  \section{Construction of Geometry of Dual Tori and SYM in
   Large $N$ Limit} \label{section_dec}

   This section is dedicated to extract geometric information
   for compactified torus which is (de)-constructed with our
   orbifold setting.

   \subsection{Target Toroidal Geometry from Orbifolding}
   \label{SUBSECTION1}


   To see how a toroidal geometry in target space arises
   from the orbifolds $\complex^3/\integer_N^2$, we
   introduce the following parametrization of $\complex^3$:
   \bea
   \label{POLAR}
    \nonumber
    z^1 &=& \rho_1 e^{i(\varphi^1 - \varphi^2
    - \varphi^3)}/\sqrt{2},\\
    \nonumber
    z^2 &=& \rho_2 e^{i(\varphi^1 + \varphi^2)}/\sqrt{2},\\
    z^3 &=& \rho_3 e^{i(\varphi^1 + \varphi^3)}/\sqrt{2}
   \eea
   where all of $\varphi^a$ run from $0$ to $2\pi$. The
   orbifold conditions in Eqs.~(\ref{ORBIFOLD}) now are
   expressed as
   \bea
   \nonumber
   \mbox{I:} &&
   \varphi^1\sim\varphi^1,
   \varphi^2\sim\varphi^2+2\pi/N,
   \varphi^3\sim\varphi^3; \\
   \mbox{II:} &&
   \varphi^1\sim\varphi^1,
   \varphi^2\sim\varphi^2,
   \varphi^3\sim\varphi^3+2\pi/N.
   \eea
   Note that the angular parametrization of $\varphi^1$
   is not unique, but the above choice in Eq.~(\ref{POLAR})
   will be convenient for our purposes.

   With the parametrization (\ref{POLAR}), the metric of
   the orbifold $\complex^3/\integer_N^2$ is
   \be\label{METRIC}
     ds^2 = \sum\limits_{a=1}^3 (d\rho_a^2
     + \rho_a^2 d \vartheta_a^2),
   \ee
   in which
   $\vartheta_1 = \varphi^1 - \varphi^2/N - \varphi^3/N$,
   $\vartheta_2 = \varphi^1 + \varphi^2/N$,
   $\vartheta_3 = \varphi^1 + \varphi^3/N$.
   If we suppress the variations of the radial coordinates and
   of $\varphi^1$, taking $\rho_a$ to be nonnegative constants
   $c_a=:Nf_a$, then the orbifold metric in Eq.~(\ref{METRIC})
   becomes
   \be\label{METRIC1}
    ds^2 = \sum\limits_{\alpha,\beta = 2}^3
       g_{\alpha\beta}d\varphi^\alpha d\varphi^\beta,
   \ee
   in which
   \be\label{METRIC2}
    (g_{\alpha\beta}) =
    \left(
     \begin{array}{cc}
      f_1^2 + f_2^2 & f_1^2\\
      f_1^2 & f_1^2 + f_3^2
     \end{array}
    \right).
   \ee

   Let us take a moment here to recall the condition(s)
   under which Eq.~(\ref{METRIC1}) gives rise to a
   legitimate Riemannian geometry, namely the metric
   $g_{\alpha\beta}$ is positive definite. From linear
   algebra this requires that $f_1^2+f_2^2>0$ and the
   determinant $g \equiv \det g_{\alpha\beta} >0$.
   Therefore, at most one of $f_a$ can vanish.

   The metric described by Eq.~(\ref{METRIC2}) is just
   that of a flat torus. (From now on we will omit the
   adjective ``flat''.) It is known that the geometry of
   a torus is specified by the complex structure modulus
   $\tau = \tau_1 + i \tau_2$ and by its area. To extract
   them, we rewrite the metric in Eq.~(\ref{METRIC2}) in
   the conformally flat form:
   \be\label{GMETRIC}
    ds^2 = e^{2\omega} |d\varphi^2 + \tau d\varphi^3|^2
   \ee
   where $e^{2\omega} = f_1^2 + f_2^2$. Then the modular
   parameter can be read off as
   \be
    \tau = e^{-2\omega}(f_1^2 + i \sqrt{g}).
   \ee
   As for the area of the torus, it is the coordinate
   area, $(2\pi)^2$, multiplied by $\sqrt{g}$:
   \be\label{AREA}
    A_{T^2} = (2\pi)^2 \sqrt{g}
    = (2\pi)^2\sqrt{f_1^2 f_2^2 + f_2^2 f_3^2 + f_3^2 f_1^2}.
   \ee

   Note that the global geometry of a two-torus can be
   described either by $g_{\alpha\beta}$ locally with
   the fixed coordinate domain (for a Euclidean
   worldsheet) or by the global characters of the area
   and the modular parameter. Thus one may use the local
   parametrization $(f_1,f_2,f_3)$ or the global one
   $(\omega,\tau_1,\tau_2)$, or even a mixed set
   $(f_1,\sqrt{g},\omega)$ to describe the geometry of
   the target torus. For example, the area $A_{T^2}$
   in Eq.~(\ref{AREA})
   can also be calculated by embedding the torus as a
   parallelogram spanned by two vectors $2\pi e^\omega$
   and $2\pi e^\omega\tau$ in a complex plane, namely
   \be\label{AREA_G}
    A_{T^2} = (2\pi e^\omega)(2\pi e^\omega \tau_2).
   \ee

   In summary, the geometry of the toroidal
   compactification of the target space is encoded
   in the limit $N\rightarrow \infty$,
   $c_a\equiv \bra\rho_a\ket \rightarrow \infty$
   with $f_a=c_a/N$ fixed.

   \subsection{World Volume Toroidal Geometry from
        (De)Construction}
   \label{SUBSECTION2}

   In BFSS Matrix theory and, subsequently, in our
   quiver matrix mechanics model (\ref{OA}), the
   target space coordinates are promoted to matrices.
   Though this increases technical complications
   to certain extent, we will see that the ideas on
   how a toroidal geometry emerges in the continuum
   limit (as a large-$N$ limit) still apply.
   Furthermore, besides the geometry for the
   compactified torus in target space, we will see
   another toroidal geometry, dual to the former,
   emerging on the world volume that is (de)constructed
   in the same limit. This is another incarnation of
   the so-called target-space/world-volume duality
   that we realized before in \cite{DW1,DW} in
   orbifolded Matrix theory.

   First let us try to implement the angular
   parametrization at the matrix level, and to
   see whether a discrete geometry can make sense
   when we assign non-vanishing vacuum expectation
   values (VEV) to the matrix counterpart of the
   variables $\rho_a$.

   As the solution to the orbifold conditions (\ref{OC}),
   the block decomposition of the bi-fundamental bosonic
   matrix variables $Z^a$ in Eq.~(\ref{OA}) can be read
   off directly from the quiver diagram Fig.~\ref{QDG}:
   \be\label{Z}
    Z^a_{mn,m^\pr n^\pr} = z^a(m,n) (\hat{V}_a)_{mn,m^\pr n^\pr},
   \ee
   in which
   \bea
    \nonumber
    (\hat{V}_2)_{mn,m^\pr n^\pr}&=&(V_N)_{m,m^\pr} \delta_{n,n^\pr}, \\
    (\hat{V}_3)_{mn,m^\pr n^\pr}&=&\delta_{m,m^\pr} (V_N)_{n,n^\pr}
   \eea
   and $\hat{V}_1:=\hat{V}_2^\dag \hat{V}_3^\dag$. Here
   the clock and shift matrices $U_N$, $V_N$ of rank $N$ are
   defined by
   \be\label{CS}
    U_N^N=\unit_N,
    V_N^N=\unit_N,
    V_NU_N=\omega_NU_NV_N,
   \ee
   with $\unit_N$ the unit matrix of rank-$N$.
   The block decomposition of other variables can be read off
   in the same way; for example,
   \be\label{Y}
    Y^i_{mn,m'n'}=y^i(m,n)\delta_{mm^\pr}\delta_{nn^\pr}.
   \ee
   At a fixed site $(m,n)$, $z^a$ (as well as $y^i$) is
   a $K$-by-$K$ matrix. To (de)construct the toroidal
   geometry, after orbifolding we need to assign nonzero
   vacuum expectation value (VEV) to each $z^a(m,n)$;
   namely, we make the following decomposition
   \be\label{DECOM}
    z^a = \bra z^a \ket + \tld{z}^a,
   \ee
   in which $\bra z^a \ket$ are the VEV and $\tld{z}^a$
   the fluctuations. We take
   \be\label{VEV}
    \bra z^a \ket \equiv \frac{f_a}{\sqrt{2}\LC} \unit_K
   \ee
   where
   \be
    \LC := 2\pi/N\,
   \ee
   $f_a$ will be understood as the same quantities that we have
   introduced in last subsection, while $\LC$ as the
   lattice constant for world volume coordinates later.

   The most direct way to look for an interpretation in
   terms of discrete geometry is to rewrite the following
   term in Eq.~(\ref{OA})
   \be\label{ACTION_YZ}
    S_{YZ} = - \int dt Tr\{|[Z^a,Y^i]|^2\},
   \ee
   as the discretized kinetic term of $Y^i$. Here we have
   absorbed $R_{11}$ into a redefinition of the world-line
   time $t^\pr=R_{11}t$ and suppressed the superscript prime.

   We will introduce the discretized derivatives by using the
   shift matrix. In this paper, the clock matrix is represented
   by $U_N = diag(\omega_N,\omega_N^2,\ldots,\omega_N^N)$
   and the shift matrix by
   \be\label{SHIFT}
    V_N = \left(
     \begin{array}{ccccc}
      0&1&0&\ldots &0\\
      0&0&1&\ldots &0\\
       & & &\ddots & \\
      0&0&0&\ldots &1\\
      1&0&0&\ldots &0
     \end{array}
    \right).
   \ee
   Note that the representation of $V_N$ in Eq.~(\ref{SHIFT})
   is the hermitian conjugate of the representation used
   in ref. \cite{DW1}. Now let $f$ be a diagonal matrix in
   the site indices:
   $f_{mn,m^\pr n^\pr} = f(m,n)\delta_{mm^\pr}\delta_{nn^\pr}$.
   The action of the shift operator, $S_a$, by a unit along
   the $a$-th direction is given by
   \be
   S_a f = \hat{V}_a f \hat{V}_a^{\dag};
   \ee
   indeed we have explicitly
   \bea
   \nonumber
    S_1 f(m,n) &=& f(m-1,n-1), \\
   \nonumber
    S_2 f(m,n) &=& f(m+1,n), \\
    S_3 f(m,n) &=& f(m,n+1).
   \eea
   Subsequently we define the discrete partial derivatives by
   \be\label{PARTIAL}
    \lp_a f := (S_a f- f)/\LC.
   \ee
   (So $\LC$ serves as a (coordinate) lattice constant.)
   Because of the relation
   \be\label{PART}
    \lp_1 = -S_2^{-1}S_3^{-1}\lp_2 - S_3^{-1}\lp_3,
   \ee
   $\lp_a$ are not algebraically independent.

   Since we will take the large $N$ limit eventually, we need to
   regularize the trace that appeared in Eqs.~(\ref{OA}) and
   (\ref{ACTION_YZ}) by
   \be\label{TRACE}
    Tr\{.\} \rightarrow \sum_{m,n} \LC^2 \kappa \{tr\{.\}\},
   \ee
   in which $\kappa$ is a regularization constant to be specified
   later, and $tr$ the trace on the subspace supporting the gauge
   group $U(K)$. Now with a little algebra, we can rewrite $S_{YZ}$
   as
   \be\label{SYZ0}
    S_{YZ} = - \int dt \sum \LC^2 \kappa tr \{
    |\LC z^a\lp_a y^i + [\tld{z}^a,y^i]|^2
    \}.
   \ee
   Below, we assume scalings that all of the variables
   including the fluctuations in Eq.~(\ref{DECOM}) are
   of $\order(1)$ in the large $N$ limit except for
   $\bra z^a\ket$ which behaves like $\order(N)$, provided
   the constants $f_a$ are independent of $N$. This is the
   common circumstance for deconstruction in the present
   literature. Separating the fluctuation field $\tld{z}^a$
   into hermitian and anti-hermitian part, $\Re\tld{z}^a$
   and $i\Im\tld{z}^a$ respectively, Eq.~(\ref{SYZ0}) can
   be further written as
   \be\label{SYZ1}
    S_{YZ} = - \int dt \sum \LC^2 \kappa tr \{
   |{f_a\over\sqrt{2}}(\lp_a y^i + {i\over f_a}
   [\sqrt{2}\Im\tld{z}^a,y^i]) +[\Re\tld{z}^a,y^i]
    +\LC\tld{z}^a\lp_a y^i|^2
    \}.
   \ee

   To reveal the discrete geometry on the quiver
   diagram, let us switch off the fluctuations in
   Eq.~(\ref{SYZ1}), resulting in
   \be\label{SYZ}
    S_{Y\bra Z\ket} = - \int dt \sum \LC^2 \kappa
    {1\over 2}  tr \{ \tld{g}^{22}(\lp_2 y^i)^2
    + \tld{g}^{33}(S_2\lp_3 y^i)^2 +
    2\tld{g}^{23}(\lp_2 y^i) (S_2\lp_3 y^i)\}.
   \ee
   Here the (contravariant) metric is defined by
   \be\label{CVMETRIC}
    (\tld{g}^{\alpha\beta})=
    \left(
     \begin{array}{cc}
      f_1^2 + f_2^2 & f_1^2\\
      f_1^2 & f_1^2 + f_3^2
     \end{array}
    \right),
   \ee
   with $\alpha, \beta = 2,3$. It is amusing to notice
   that comparing with Eq.~(\ref{METRIC2}), we have
   \be\label{DUAL}
    \tld{g}^{\alpha\beta} = g_{\alpha\beta}.
   \ee

   As a corollary of either Eq.~(\ref{CVMETRIC}) or
   Eq.~(\ref{DUAL}), we see that $\tld{g}^{\alpha\beta}$
   is independent of $N$! As a simplest application of
   the metric (\ref{CVMETRIC}) in discrete geometry, we
   can assign an area to the elementary parallelogram, or
   {\em plaque} in the jargon of lattice gauge theory,
   spanned by two edges labelled by $Z^2$ and $Z^3$
   in the quiver diagram in Fig.~\ref{QDG}:
   \be\label{EAREA}
    \delta A_N = {\LC^2\over\sqrt{g}}.
   \ee
   where $\LC^2$ is the coordinate area. Because of
   translation invariance, by counting the total number
   of the plaques we get a total area for the quiver
   diagram:
   \be\label{AN}
    A_N = N^2 \cdot \delta A_N = {(2\pi)^2\over\sqrt{g}},
   \ee
   which is also independent of $N$.

   In the continuum limit, i.e. in the large $N$ limit
   with $f_a$ fixed, we have
   \be\label{CLTORUS}
    (\integer_N^2, \LC^2)
    \rightarrow (\tld{T}^2, d\sigma^2 d\sigma^3);
   \ee
   namely the quiver diagram (de)constructs a
   continuum torus $\tld{T}^2$, with continuous
   coordinates $\sigma^\alpha$ ($\alpha=2,3$)
   running from $0$ to $2\pi$. So the metric in Eq.~(\ref{CVMETRIC})
   on the discrete quiver diagram
   survives the large $N$ limit, and becomes the
   metric on the torus $\tld{T}^2$.

   Moreover, in the large $N$ limit,
   \be
    \lp_2\rightarrow \partial/\partial\sigma^2,
    \lp_3\rightarrow \partial/\partial\sigma^3,
   \ee
   and because of the linear relation in Eq.~(\ref{PART})
   \be\label{P1}
     \lp_1\rightarrow -\partial/\partial\sigma^2
     -\partial/\partial\sigma^3.
   \ee
   Now it is easy to work out the large $N$ limit of
   $S_{Y\bra Z\ket}$:
   \be\label{CYZ}
    S_{Y\bra Z\ket} = -\int\limits dtd^2\sigma\kappa tr\{{1\over 2}
    \tld{g}^{\alpha\beta}\partial_\alpha y^i\partial_\beta y^i
    \}\, ,
   \ee
   where, as usual, $d^2\sigma = d\sigma^2 d\sigma^3$.
   The same positive-definiteness condition analyzed below
   Eq.~(\ref{METRIC2}) should be imposed to ensure the
   positive-definiteness of $\tld{g}^{\alpha\beta}$. It
   is a constraint on VEV in Eq.~(\ref{VEV}), which gives
   rise to a normal Riemannian geometry to the toroidal
   membrane, $\tld{T}^2$, (de)constructed with our quiver
   diagram.

   To see that the toroidal geometry $\tld{T}^2$ is dual to
   that of the compactified target torus $T^2$, we choose
   the regularization constant $\kappa$ in Eq.~(\ref{TRACE})
   to be $\kappa=1/\sqrt{g}=\sqrt{\tld{g}}$, where $\tld{g}$
   is the determinant of $\tld{g}_{\alpha\beta}$, the inverse
   of $\tld{g}^{\alpha\beta}$ in Eq.~(\ref{CVMETRIC}),
   such that $d^2\sigma \cdot \kappa$ becomes the invariant
   measure on the dual torus. Here we introduce the definition
   that two tori
   $(T^2,(\varphi^2,\varphi^3),g_{\alpha\beta})$ and
   $(\tld{T}^2,(\sigma^2,\sigma^3),\tld{g}_{\alpha\beta})$
   are dual to each other, if and if all of the affine
   parameters have the same domain from $0$ to $2\pi$
   and Eq.~(\ref{DUAL}) is satisfied. Consequently, the
   area of the dual torus is
   \be\label{DUALAREA}
    A_{\tld{T}^2} = \int d^2\sigma\kappa
    = {(2\pi)^2\over \sqrt{g}},
   \ee
   which coincides with $A_N$ in Eq.~(\ref{AN}). Just as
   the case for $A_{T^2}$ in  Eq.~(\ref{AREA_G}), the
   result for $A_{\tld{T}^2}$ in Eq.~(\ref{DUALAREA}) can
   also be obtained by rewriting the dual metric as
   \be
    d\tld{s}^2 = e^{2\Omega}
     |d\sigma^2 + \tld{\tau}d\sigma^3|^2,
   \ee
   in which $e^{2\Omega} = (f_1^2+f_3^2)/g$, and the dual
   modular parameter is identified to be
   \be
      \tld{\tau} = \tld{\tau}_1 + i \tld{\tau}_2
      = \frac{-f_1^2 + i\sqrt{g}}{e^{2\Omega} g}
      = - {1\over \tau}!
   \ee

   Thus, Eq. (\ref{CYZ}) can be written as
   \be\label{CYZ2}
     S_{Y\bra Z\ket} = -\int\limits dtd^2\sigma
     \sqrt{\tld{g}} \tld{g}^{\alpha\beta} tr\{{1\over 2}
     \partial_\alpha y^i\partial_\beta y^i \}\, ,
   \ee
   Without any additional pain we can safely claim that
   in Eq.~(\ref{OA}),
   \be\label{Y4}
    S_{Y^4}:=\int dt Tr\{{1\over 4}[Y^i,Y^j]^2\}
           \stackrel{N\rightarrow\infty}{\longrightarrow}
    S_{Y^4} = \int\limits dtd^2\sigma \sqrt{\tld{g}}
           tr\{{1\over 4} [y^i,y^j]^2\}.
   \ee

   To summarize, the above terms are of the usual form of
   the action integral for fields on the world volume of
   a torus, with $\tld{g}^{\alpha\beta}$ as
   contravariant metric. Later we will see that in the
   continuum limit, all other terms in our quiver model
   contain the same metric. This feature indeed identifies
   $\tld{g}^{\alpha\beta}$ as the metric on the world volume
   (de)constructed by our orbifolds. In last subsection, we
   drew the quiver diagram in Fig.~\ref{QDG} as a square
   lattice; however, no notion of length was introduced at
   that stage. It was only after assigning non-zero VEV as
   in Eq.~(\ref{VEV}), the quiver diagram becomes a lattice
   with meaningful lattice constant, and in the large $N$
   limit becomes a continuum torus, $\tld{T}^2$, with a
   flat metric. The relation (\ref{DUAL}) implies that the
   toroidal geometry of $\tld{T}^2$ is {\it dual to} that
   of the compactified torus, $T^2$, in target space as we
   discussed in last subsection. In this way, our
   (de)construction procedure (orbifolding, assigning non-zero
   VEV and taking the continuum limit) exhibits the so-called
   target-space/world-volume duality. In the literature,
   including our previous paper \cite{DW1}, this duality was
   shown only for regular tori; here we have shown the validity
   of this duality when the compactified target torus is of
   a generic (oblique) shape.

   \subsection{$1+2$-Dimensional Super Yang-Mills and the
       Detour into Four Dimensions}

   We devote this subsection to a complete discussion of
   the continuum limit of our orbifolded quiver matrix
   mechanics. On the one hand, we will show that all terms
   in the continuum action contain one and same metric
   $\tld{g}^{\alpha\beta}$, justifying the emergence of
   the world volume geometry in (de)construction through
   orbifolding. On the other hand, we will show that in
   this continuum limit, the quiver matrix mechanics
   approaches to $1+2$-dimensional SYM on $\tld{T}^2$.
   Previously to argue for the $S$-duality and rotational
   invariance in Matrix Theory compactified on a torus, a
   connection between $1+2$ and $1+3$ dimensional SYM was
   proposed in refs. \cite{S,SS}. In this subsection we
   will see that indeed this detour into four dimensions
   is something very natural in the present approach.

   In \cite{DW1} we have studied the case with $f_1=0$,
   leading to a regular torus. To consider torus of more
   general shape, here we study another simplified case,
   corresponding to a triangular lattice, with
   \be\label{SVEV}
    f_1=f_2=f_3=L.
   \ee
   In accordance with the parametrization (\ref{POLAR})
   and the VEV in Eqs.~(\ref{VEV}) and (\ref{SVEV}), we
   parameterize fluctuations in Eq.~(\ref{DECOM}) by
   \bea
   \label{FLUC}
    \tld{z}^1 &=& (\phi^1 + iL(\phi_1^\pr-A_2-A_3))/\sqrt{2},
    \nonumber \\
    \tld{z}^2 &=& (\phi^2 + iL(\phi_1^\pr+A_2))/\sqrt{2},
    \nonumber \\
    \tld{z}^3 &=& (\phi^3 + iL(\phi_1^\pr+A_3))/\sqrt{2}.
   \eea
   All the new variables here, with the site indices
   $(m,n)$ omitted, are $K$-by-$K$ matrices. In the
   following we will discuss the dynamics of these
   fluctuations.

   \subsubsection{Discrete Geometry and Equilateral
      (Triangular) Lattice}

\begin{figure}[hbtp]
    \centering
    \psfrag{l1}[][]{$\lambda^{100}$}
    \psfrag{l2}[][]{$\lambda^{010}$}
    \psfrag{l3}[][]{$\lambda^{001}$}
    \psfrag{l0}[][]{$\lambda^{000}$}
    \psfrag{z1}[][]{$z^1$}
    \psfrag{z2}[][]{$z^2$}
    \psfrag{z3}[][]{$z^3$}
    \psfrag{z0}[][]{$A_0,y^{1,2,3}$}
    \includegraphics[width=0.70\textwidth]{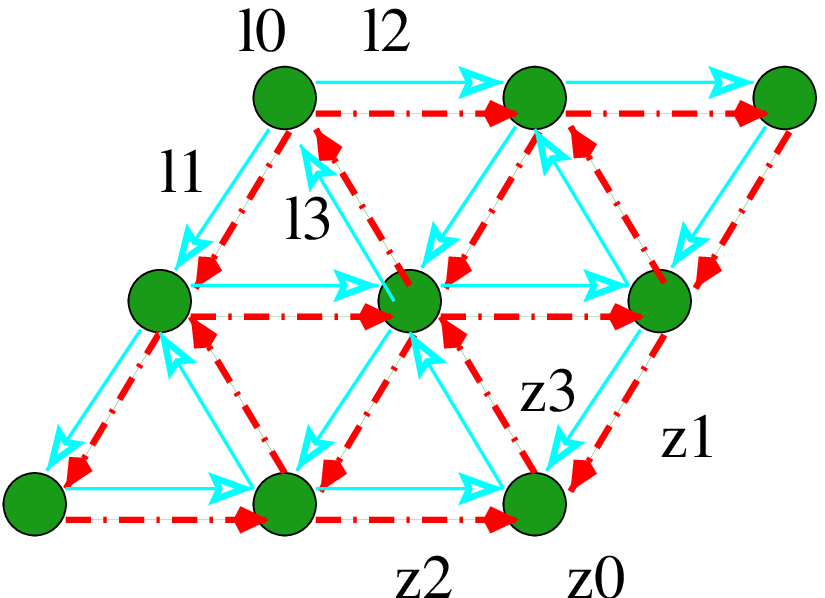}
    \caption{Two-dimensional Equilateral Lattice from
        (De)Construction}
    \label{QDG2}
   \end{figure}

   With the symmetric VEV (\ref{SVEV}), the quiver
   diagram in Fig.~\ref{QDG} becomes a equilateral
   triangular lattice,  shown in Fig.~\ref{QDG2},
   in which re-label the fermionic coordinates
   \bea
   \nonumber
    \Lambda^{s_0s_1s_2s_3}
     =\lambda^{s_0s_1s_2s_3} \hat{V}_{s_1+2s_2+3s_3},
        &\mbox{for}& s_1+s_2+s_3=0,1,\\
      \Lambda^{s_0s_1s_2s_3} =\lambda^{s_0s_1s_2s_3}
      \hat{V}_{(s_1+1)+2(s_2+1)+3(s_3+1)}^\dag
        &\mbox{for}& s_1+s_2+s_3=2,3
   \label{LV}
   \eea
   with $s_c+1$ (with $c=1,2,3$) defined modulo $2$.

   In fact, with the VEV (\ref{SVEV}), the
   metric in Eq.~(\ref{CVMETRIC}) becomes
   \be\label{SMETRIC}
    (\tld{g}^{\alpha\beta})
    =L^2
    \left(
     \begin{array}{cc}
      2&1\\
      1&2
     \end{array}
    \right)
   \ee
   whose inverse is
   \be
   \label{CV_METRIC}
    (\tld{g}_{\alpha\beta})={1\over 3L^2}
    \left(
     \begin{array}{cc}
      2&-1\\
      -1&2
     \end{array}
    \right).
   \ee
   From either the definition of the discrete partial derivatives
   in Eq.~(\ref{PARTIAL}) or the large $N$ limit of the measure
   on $\integer_N^2$ in Eq.~(\ref{CLTORUS}), we learn that the
   role of the coordinate lattice constant in both two directions
   on $\integer_N^2$ is played by $\LC$, so that the total
   coordinate length of each cycles in either discrete or
   continuum cases is just $2\pi$. Now we compute the {\em
   proper length} of the unit vectors $e_a$, denoted as
   $||e_a||$, in three directions in Fig.~\ref{QDG2} labelled
   by $z^a$, with
   $e_1=\LC\cdot(-1,-1)^T$,
   $e_2=\LC\cdot(1,0)^T$,
   $e_3=\LC\cdot(0,1)^T$.
   The result is
   \be\label{LENGTH}
       ||e_a||^2 = (e_a,e_a)
    = \tld{g}_{\alpha\beta}e_a^\alpha e_a^\beta
    = {2\LC^2\over 3L^2},
   \ee
   for all $a=1,2,3$ (no summation on $a$). Therefore
   the lattice is equilateral, with the area
   $A_{\tld{T}^2}= {(2\pi)^2 /\sqrt{3}L^2}$
   from Eq.~(\ref{DUALAREA}).

  A linear transformation
   \be
   \label{LTRANS}
    \left(\begin{array}{c}
     \sigma^2\\\sigma^3
     \end{array}
    \right)={L\over\sqrt{2}}
    \left(\begin{array}{cc}
     \sqrt{3}&1\\
     0&2
     \end{array}
    \right)
    \left(\begin{array}{c}
     w^1\\w^2
     \end{array}
    \right),
    \left(\begin{array}{c}
     w^1\\w^2
     \end{array}
    \right)={1\over\sqrt{6}L}
    \left(\begin{array}{cc}
     2&-1\\
     0&\sqrt{3}
     \end{array}
    \right)
    \left(\begin{array}{c}
     \sigma^2\\\sigma^3
     \end{array}
    \right),
   \ee
   transforms the dual metric into the standard form
   \be\label{E}
    d\tld{s}^2 = (dw^1)^2 + (dw^2)^2,
   \ee
   with the measure $d^2\sigma/\sqrt{3}L^2 = d^2w$. We can
   calculate the area $A_{\tld{T}^2}$ in the $w$-frame.
   In fact, $\tld{T}^2$ in $\sigma$-frame spanned by two
   basis vectors
   $E_2:=2\pi\cdot (1,0)^T$, $E_3:=2\pi\cdot (0,1)^T$;
   by the second formula in Eq.~(\ref{LTRANS}),
   in $w$-frame,
   \be
    E_2 = {\sqrt{2}\over \sqrt{3}L}\cdot 2\pi\cdot (1,0)^T,
    E_3 = {\sqrt{2}\over \sqrt{3}L}\cdot 2\pi\cdot
  (-{1\over 2},{\sqrt{3}\over 2})^T.
   \ee
   The value of $A_{\tld{T}^2}$ follows hence because of Eq.~(\ref{E}).


   \subsubsection{Four Dimensions}

   Now it is the time to revisit $S_{YZ}$ in Eq.~(\ref{SYZ1}).
   Recall the term $\LC \tld{z}^a\lp_a y^i$ is of order
   $\order(1/N)$, it is not difficult to deduce the continuum
   limit of $S_{YZ}$;
   \be
   \label{YGK1}
    S_{YZ} = - \int dt d^2\sigma {1\over \sqrt{3}L^2} {1\over 2}
    tr \{ L^2(D_a y^i+i[\phi_1^\pr, y^i])^2 - [\phi^a, y^i]^2\}
   \ee
   where the index $a$ runs from $1$ to $3$, and
   $D_\alpha = \partial_\alpha + i[A_\alpha,.]$
   for $\alpha = 2,3$. Recall the fact that
   \be\label{SD}
    D_1+D_2+D_3=0;
   \ee
   we get
   \be
   \label{YGK2}
   S_{YZ} = - \int dt d^2\sigma {1\over \sqrt{3}L^2} {1\over 2}
   tr \{ \tld{g}^{\alpha\beta}D_\alpha y^i D_\beta y^i
   - 3L^2[\phi_1^\pr, y^i]^2 -   [\phi^a, y^i]^2\}.
   \ee

   Eq.~(\ref{YGK2}) is of a standard form in SYM after a
   rescaling of $\phi_1^\pr$. More amusing is the assertion
   that the continuum action Eq.~(\ref{YGK1}) can be obtained
   from dimension reduction from SYM in one more dimension!
   Our key observation here is that the role of $\phi_1^\pr$
   in Eqs.~(\ref{FLUC}) and (\ref{YGK1}) is very similar
   to a gauge connection, whose direction can be parameterized
   virtually by a coordinate $\sigma^1$. This implies that
   $S_{YZ}$ in Eq.~(\ref{YGK1}) can be regarded as a
   $1+3$-dimensional theory subject to the dimensional
   reduction constraint
   \be\label{DR}
    \partial/\partial\sigma^1 \equiv 0.
   \ee
   This motivates us to introduce a three-dimensional
   Euclidean space $(x^1,x^2,x^3)$,
   in the sense of a covering space,
   such that
   \bea
   \nonumber
    \xi \frac{\partial}{\partial x^1} &=&
    \eta\frac{\partial}{\partial\sigma^1}
   - \frac{\partial}{\partial\sigma^2}
     -\frac{\partial}{\partial\sigma^3},\\
   \nonumber
    \xi \frac{\partial}{\partial x^2} &=&
    \eta\frac{\partial}{\partial\sigma^1}
  + \frac{\partial}{\partial\sigma^2},\\
   \label{TRANSXS}
    \xi \frac{\partial}{\partial x^3} &=&
    \eta\frac{\partial}{\partial\sigma^1}
  + \frac{\partial}{\partial\sigma^3}
   \eea
   with two constants $\xi$, $\eta$, and $\phi_1^\pr = \eta A_1$
   with $A_1$ the gauge connection in the $\sigma^1$-direction.
   Since covariant derivatives transform in the same way as ordinary
   derivatives, from Eq.~(\ref{TRANSXS}) we can solve the coordinate
   transformation
   \be\label{TRANS}
    \left(
     \begin{array}{c}
      \sigma^1\\\sigma^2\\\sigma^3
     \end{array}
    \right) = \xi^{-1}
    \left(
     \begin{array}{ccc}
      \eta&\eta&\eta\\
      -1&1&0\\
      -1&0&1
     \end{array}
    \right)
    \left(
     \begin{array}{c}
      x^1\\x^2\\x^3
     \end{array}
    \right).
   \ee

   Now we suppose the terms $[\phi^a,y^i]^2$ in Eq.~(\ref{YGK1}) are
   intact in dimensional reduction, and the metric in the $x$-frame
   is the standard Euclidean metric. Then we have to modify the
   regularization of the trace in Eq.~(\ref{TRACE}) and interpret
   the measure in Eq.~(\ref{YGK1}) in the following way
   \be\label{MEASURE}
    Tr\rightarrow \int\int {d^2\sigma\over \sqrt{3}L^2}\kappa^\pr =
    \int\int d\sigma^2 d\sigma^3 \int d\sigma^1 \sqrt{G^\sigma}
    =\int\int\int d^3x \, ,
   \ee
   where we have written the multiple integral explicitly, and
   $\kappa^\pr$ is the proper length of the one-dimensional space
   swept by $\sigma^1$ and $G^\sigma$ is the determinant of the
   covariant metric in the three-dimensional $\sigma$-frame.
   Keeping Eq.~(\ref{MEASURE}) in mind and naming the covariant
   derivatives in $x$-frame as $\nabla_i$ for $i=1,2,3$,
   the first terms in Eq.~(\ref{YGK1}) must be of the standard
   form $\nabla_j y^i\nabla_j y^i$, from which $\xi$ is fixed
   to be $1/L$.

   Consequently, we can write the covariant metric in
   $\sigma$-frame as
   \be\label{3CMETRIC}
    (G^{aa^\pr}_\sigma) = L^2
    \left(
     \begin{array}{ccc}
      3\eta^2&0&0\\0&2&1\\0&1&2
     \end{array}
    \right)
   \ee
   whose restriction to the lower-right $2$-by-$2$ block is
   identical to $\tld{g}^{\alpha\beta}$ in Eq.~(\ref{SMETRIC});
   similarly for the contravariant metric
   \be\label{3SMETRIC}
    G^\sigma_{aa^\pr} ={1\over 3L^2}
    \left(
     \begin{array}{ccc}
      {1\over \eta^2} &0&0\\0&2&-1\\0&-1&2
     \end{array}
    \right).
   \ee
   Note that $\eta$ is a free parameter at this stage because
   only $\phi_1^\pr$ is ``visible" after dimensional reduction.
   To make Eqs.~(\ref{TRANSXS}) and (\ref{TRANS}) nonsingular,
   we only require $\eta$ nonvanishing.
   In fact, $\eta$ controls the scale of the basis vector in
   $\sigma^1$-direction, whose magnitude is not relevant in
   the present context. For later convenience we rescale
   $\sigma^1\rightarrow \sigma^1/\sqrt{3}L\eta$ such that
   $G^\sigma_{11} \rightarrow 1$ and that the domain of
   $\sigma^1$ changes to $(0,\kappa^\pr)$; now
   $\eta$ is entirely absorbed.

   From Eq.~(\ref{TRANSXS}), one can easily infer that viewed in
   $x$-frame, $\sigma^2$ runs along the $(-1,2,-1)$-direction,
   $\sigma^3$ in the $(-1,-1,2)$-direction, while $\sigma^1$ in
   the $(1,1,1)$-direction. The geometric significance of the
   dimensional reduction condition
   $\partial/\partial\sigma^1\equiv 0$ is nothing but the
   requirement the directional derivative in the
   $(1,1,1)$-direction in $x$-frame should vanish, i.e.
   to restrict the theory to the sector invariant under
   translations in $(1,1,1)$-direction. (Of course,
   on $(\sigma^2,\sigma^3)$-plane, we need to impose
   periodic boundary conditions to make it into the
   torus $\tld{T}^2$.)

   As a final remark in this sub-subsection,
   we see that $\kappa^\pr$ provides the room for the
   electric-magnetic duality argument in \cite{SS}
   to convince $O(8)$ rotational symmetry in IIB string
   theory. We know that the Matrix Theory compactified
   on a three-dimensional torus is equivalent to
   $1+3$-dimensional SYM. In fact, there exist two
   different limits from this $1+3$-dimensional SYM
   to IIB string theory, in which $\kappa^\pr$ are
   treated differently. The first is in the sense of
   Sethi and Susskind in \cite{SS}, that the $\kappa^\pr$
   is tuned to be proportional to the overall size of
   $\tld{T}^2$. Then in the second one, the
   $\sigma^1$-direction is taken as a KK circle,
   equivalently to decompactify the dual circle in
   the target space, and the original $1+2$-dimensional theory
   is in the KK limit; in this case, there appears an
   additional wave function normalization such that
   effectively $\kappa^\pr = 1$ (the dimensional
   reduction condition (\ref{DR}) is equivalent to
   the prescription to keep only the zero-modes along
   the KK-circle).

   \subsubsection{SYM in the Continuum Limit}
   \label{SYMC}

   With the above preparations, now it is straightforward
   to derive the continuum limit of the quiver matrix
   mechanics (\ref{OA}) and show that the full action
   is none other than a $1+2$-dimensional SYM with $16$
   supercharges. We will bring the power of the
   KK dimensional reduction of $1+3$-dimensional SYM
   into full play, with $\kappa^\pr$ taken to be unity.

   \noindent
   {\bf I. Scalar}

   Collect Eqs.~(\ref{Y4}), (\ref{YGK2}) and the kinetic
   term of $Y^i$ (the first term) in Eq.~(\ref{OA}).
   \be\label{3Y}
    S_Y = \int dt {d^2\sigma \over \sqrt{3}L^2} tr\{
      {1\over 2} ((D_t y^i)^2
    - \tld{g}^{\alpha\beta} D_\alpha y^i D_\beta y^i)
    + {1\over 2}[\phi^A, y^i]^2 + {1\over 4} [y^i,y^j]^2
    \}
   \ee
   where the index $A$ runs from $1$ to $4$ and
   $\phi^4:=\sqrt{3}L\phi_1^\pr$; or in a mixed fashion,
   with three-dimensional measure and four-dimensional
   Lagrangian:
   \be\label{4Y}
    S_Y = \int dt {d^2\sigma \over \sqrt{3}L^2} tr\{
    {1\over 2} ((D_t y^i)^2 - (\nabla_j y^i)^2)
    + {1\over 2}[\phi^a, y^i]^2 + {1\over 4} [y^i,y^j]^2
    \}
   \ee

   \noindent
   {\bf II. Yang-Mills}

   For the bosonic bi-fundamental variables $Z^a$,
   we rewrite the relevant terms in Eq.~(\ref{OA})
   \be
   \label{SZ}
    S_Z = \int dt Tr \{
    |[D_t, Z^a]|^2
    - \frac{1}{2}(|[Z^a,Z^{a^\pr \dag}]|^2
    + |[Z^a,Z^{a^\pr}]|^2)\}.
   \ee
   Introduce $[Z^a, Z^{a^\pr\dag}] =:
     \hat{V}_{a^\pr}^\dag P_{a^\pr a} \hat{V}_a$
   such that $P_{a^\pr a}^\dag = P_{aa^\pr}$ and
   \be
    P_{a^\pr a}
    = S_{a^\pr}z^a S_a z^{a^\pr\dag} - z^{a^\pr\dag}z^a
    =
    \LC \lp _{a^\pr} z^a S_a z^{a^\pr \dag}
    + \LC z^a \lp_a z^{a^\pr \dag}
    + [z^a,z^{a^\pr\dag}].
   \ee
   Then $[Z^a,Z^{a^\pr}] =: Q_{aa^\pr} \hat{V}_a\hat{V}_{a^\pr}$
   such that $Q_{a^\pr a} = - Q_{aa^\pr}$ and
   \be
    Q_{aa^\pr} = z^aS_az^{a^\pr} - z^{a^\pr}S_{a^\pr}z^a
    =
    \LC z^a\lp_a z^{a^\pr}
    -\LC z^{a^\pr}\lp_{a^\pr}z^a + [z^a,z^{a^\pr}].
   \ee
   Also $[D_t,Z^a] =: S_{0a}\hat{V}_a$ with
   \be
    S_{0a} = \dot{z}^a - i\LC z^a\lp_a A_0 + i [A_0,z^a].
   \ee
   The action $S_Z$ in Eq.~(\ref{SZ}) is recast into
   \be\label{SZ1}
    S_Z = \int dt \sum\LC^2 \kappa tr \{
    |S_{0a}|^2
    - {1\over 2}(|P_{a^\pr a}|^2 + |Q_{a^\pr a}|^2)\}.
   \ee

   Now we separate the VEV and the fluctuation as in
   Eq.~(\ref{DECOM}), impose the VEV (\ref{SVEV})
   which, at large $N$, is the moduli condition in
   the language of quantum field theory, and substitute
   the parametrization of fluctuation in Eq.~(\ref{FLUC});
   remember the remark below Eq.~(\ref{TRANSXS}) and
   definition of $\nabla_i$ above Eq.~(\ref{3CMETRIC}),
   with subscripts $a$ and $a^\pr$ in the sense of the
   directions in the quiver diagram replaced by $j$ and
   $j^\pr$ in three-dimensional Euclidean space or
   three-torus, what follows in the continuum limit is
   \bea
   \label{PJJ}
    P_{j^\pr j} &=& {1\over 2} (
    (\nabla_j\phi^{j^\pr} + \nabla_{j^\pr}\phi^i) +
    i (F_{j^\pr j} -i [\phi^j,\phi^{j^\pr}])),\\
   \label{QJJ}
    Q_{jj^\pr} &=& {1\over 2} (
    (\nabla_j\phi^{j^\pr} - \nabla_{j^\pr}\phi^j) +
    i (-F_{j^\pr j} -i [\phi^j,\phi^{j^\pr}])),\\
   \label{S0J}
    S_{0j} &=& {1\over \sqrt{2}}(
    D_t\phi^j + iF_{0j}),
   \eea
   in which the gauge field strength $F_{jj^\pr}$,
   $F_{0j}$ are defined conventionally and
   each operator in Eqs.~(\ref{PJJ}), (\ref{QJJ})
   and (\ref{S0J}) is sorted in the form
   $F = \Re F + i \Im F$ so that
   $|F|^2 = (\Re F)^2 + (\Im F)^2$. Then similar to
   Eq.~(\ref{4Y}), Eq.~(\ref{SZ1}) can be put into
   \be\label{SZC}
    S_Z = \int dt \frac{d^2\sigma}{\sqrt{3}L^2} tr\{
    {1\over 2} F_{0j}^2 - {1\over 4} F_{j^\pr j}^2
    + {1\over 2} ((D_t\phi^a)^2-(\nabla_{j}\phi^a)^2)
    + {1\over 4} [\phi^a,\phi^{a^\pr}]^2 \}
   \ee
   where we recover the $a$, $a^\pr$ indices for
   scalar fields. Eq.~(\ref{SZC}) should be understood
   with the help of the dimensional reduction condition
   (\ref{DR}); however we are not bothered with deducing
   a similar expression like Eq.~(\ref{3Y}), since the
   goal here is just to check the $1+2$-dimensional SYM
   in the continuum limit. It is easy to see that
   Eq.~(\ref{4Y}) plus Eq.~(\ref{SZC}) are simply a
   dimensional reduction of the 4-dimensional Yang-Mills
   theory subjected to Eq.~(\ref{DR}).

   \noindent
   {\bf III. Fermions}

   Once again, we try to deduce the continuum limit
   for fermions from
   \be
    S_F = \int dt Tr \{
      - {i\over 2}\Lambda^{\dag}[D_t,\Lambda]
      + \frac{1}{2} \Lambda^{\dag} \gamma_i [Y^i,\Lambda]
      + \frac{1}{\sqrt{2}}\Lambda^{\dag} (\tld{\gamma}_a
    [Z^a,\Lambda] + \tld{\gamma}_a^{\dag} [Z^{a\dag},\Lambda] )
     \}
   \label{FA}
   \ee
   in the four-dimensional point of view.

   This time, we do it in the fastest way. From our experience
   with $S_Y$ and $S_Z$, we know that the effects of $\hat{V}_a$
   in Eqs.~(\ref{LV}) are washed out in the large $N$ limit
   except the terms containing the VEV of $Z^a$. We only emphasize
   that we can take the VEV's to be nonnegative in Eq.~(\ref{VEV})
   because their phases can be absorbed into the redefinition of
   the fermionic coordinates. Accordingly, we can write down the
   continuum limit of $S_F$ directly
   \be
    S_F = \int dt{d^2\sigma\over \sqrt{3}L^2}{1\over 2} tr \{
     - i\lambda^{\dag}D_t\lambda
     + \lambda^{\dag}\gamma^{2j+3} [\nabla_j,\lambda]
     + \lambda^{\dag} \gamma_i [y^i,\lambda]
     + \lambda^{\dag} \gamma_{2a+2} [\phi^a,\lambda]
    \}
   \label{FAC}
   \ee
   where $\gamma_{2a+2} = \tld{\gamma}_a + \tld{\gamma}_a^\dag$,
   $\gamma_{2j+3} = i(\tld{\gamma}_j - \tld{\gamma}_j^\dag )$.

   In summary, we have shown that the continuum limit of the
   quiver matrix mechanics in Eq.~(\ref{OA}) is a $d=1+2$ SYM:
   Eqs.~(\ref{4Y}), (\ref{SZC}) plus Eq.~(\ref{FAC}) constitute
   precisely the action of the $d=1+3$ SYM with 16 supercharges
   dimensionally reduced by Eq.~(\ref{DR}). This outcome in
   the continuum limit justifies our creed to approximate a
   compactification in Matrix Theory via a sequence of orbifolds,
   demonstrating that we are on the right track for IIB/M duality.

  \section{Wrapping Matrix Membrane and $SL(2,\integer)$ Duality}
  \label{section_mem}

   It is generally believed that in M-theory framework, IIB
   string theory can be described by M-theory compactified
   on a torus. In \cite{SCH2}, Schwarz suggested that
   in such a theory there exist solitonic states, describing
   M2-branes wrapping on the target torus, which correspond
   to the doubly charged $(q_1,q_2)$-strings (or bound FD
   strings) in IIB theory; and it is very tempting to identify
   the $SL(2,\integer)$ duality in IIB string theory with the
   geometric $SL(2,\integer)$ invariance for a torus. He
   already noted a serious problem in this identification,
   i.e. the degeneracy of wrapping membrane states would be
   generally greater than that of $(q_1,q_2)$ strings, unless
   there is a way to identify the degenerate wrapping membrane
   states which he assumed is true. However, the explicit
   description of these solitonic wrapping membranes and the
   details of how the elimination of their degeneracy happens
   are still in demand in the literature.

   Since our quiver matrix mechanics (de)constructs M-theory
   compactified on a torus, with a toroidal geometry for
   the dynamical membrane (see Subsections \ref{SUBSECTION1}
   and \ref{SUBSECTION2}), it provides a natural platform
   for dealing with the above mentioned problems involving
   wrapping membranes on target torus in IIB/M(atrix) theory.
   In a previous work \cite{DW} we have resolved these problems
   for the case when the compactified target torus is a regular
   one. In this section we generalize the discussions for an
   oblique torus with a generic complex modular parameter $\tau$.

   \subsection{Wrapping a Membrane on Orbifold}
     \label{WRAPPING}

   How to define the states with a definite wrapping number for
   wrapping matrix membranes?

   For a continuous membrane of toroidal topology, we use
   a pair of real coordinates $(q,p)$, with the equivalence
   $q\sim q+2\pi$, $p\sim p+2\pi$. The continuous wrapping
   map from the membrane to the target $T^2$ satisfies
   the periodic boundary conditions
   \bea
    \nonumber
    \varphi^2(q+2\pi,p) = \varphi^2(q,p)+2\pi m^2 &,&
    \varphi^2(q,p+2\pi) = \varphi^2(q,p)+2\pi n^2,\\
    \label{BDT}
    \varphi^3(q+2\pi,p) = \varphi^3(q,p)+2\pi m^3 &,&
    \varphi^3(q,p+2\pi) = \varphi^3(q,p)+2\pi n^3,
   \eea
   for four arbitrary integers $m^2$, $n^2$, $m^3$, $n^3$.
   The solution, up to homotopy and large diffeomorphisms,
   is of the form
   \be\label{MAPT}
    \left(
     \begin{array}{c}
      \varphi^2 (q,p) \\ \varphi^3 (q,p)
     \end{array}
    \right) =
    W(\vec{m},\vec{n})
    \left(\begin{array}{c}
     q \\ p
    \end{array}\right)
   \ee
   with the wrapping map matrix
   \be\label{MAPW}
    W(\vec{m},\vec{n}) :=(\vec{m},\vec{n}),
    \vec{m} := (m^2,m^3)^T,
    \vec{n} := (n^2,n^3)^T.
   \ee

   Subsequently the pull-back geometry from $T^2$ to
   the membrane is
   \be
    ds^2 = e^{2\omega^\star}|dq + \tau^\star dp|^2.
   \ee
   here the pull-back conformal factor
   is $e^{\omega^\star} = e^\omega |m^2 + m^3\tau|$,
   with the pull-back modular parameter to be
   \be\label{TAU}
    \tau^\star = \frac{n^3\tau + n^2}{m^3\tau + m^2}.
   \ee
   ($\tau$ is the modular parameter of $T^2$.)  So the
   induced measure is
   \be\label{IMEASURE}
     dqdp \cdot e^{2\omega^\star}\tau_2^\star
     = dqdp \cdot e^{2\omega} \cdot \Im ((n^3\tau + n^2)
(m^3\bar{\tau} + m^2)) = dqdp \cdot w \cdot \sqrt{g},
   \ee
   or simply $ d^2 \varphi = w dqdp$, with $w=detW$
   the {\em wrapping number}.
   \footnote{ Some authors gave alike construction in different
   circumstances; for example, in \cite{Bars}, Bars considered
   the connections between discrete area preserving
   diffeomorphisms, reduced Yang-Mills and strings.}

   The wrapping map (\ref{MAPT}) is fully characterized by the
   induced structure ~(\ref{TAU}) and (\ref{IMEASURE}). We know
   that, in the form of gauge-fixed metric (\ref{GMETRIC}), a
   torus possesses an $SL(2,\integer)$ symmetry containing all
   large diffeomorphisms:
   \be\label{SL2Z}
    \tau \rightarrow \frac{ a\tau + b }{ c\tau + d },
    \qquad \left(\begin{array}{c}
     \varphi^2 \\ \varphi^3
    \end{array}\right )
    \rightarrow
    C\cdot
    \left(\begin{array}{c}
     \varphi^2 \\ \varphi^3
    \end{array}\right ) \, ,
   \ee
   where $C$ is an $SL(2,\integer)$ matrix
   \be\label{SL2ZE}
    C =
    \left(
     \begin{array}{cc}
      a & -b \\
      -c & d
     \end{array}
    \right).
   \ee
   Because $\det C=1$, none of  $d^2\varphi$, $\sqrt{g}$ and
   the wrapping number $w$ is changed under this $SL(2,\integer)$.

   \subsection{Matrix States of Wrapping Membrane and Fractional Powers}

   The investigation in the previous subsection is carried out solely
   for the continuous torus. By (de)constructing a torus with a sequence
   of orbifolds, the wrapping map in Eqs.~(\ref{BDT}) change to be the
   following form
   \bea
   \nonumber
    z^1(q+2\pi,p) = e^{-i2\pi (m^2+m^3)/N}z^1(q,p)&,&
    z^1(q,p+2\pi) = e^{-i2\pi (n^2+n^3)/N}z^1(q,p),\\
   \nonumber
    z^2(q+2\pi,p) = e^{i2\pi m^2/N}z^2(q,p)&,&
    z^2(q,p+2\pi) = e^{i2\pi n^2/N}z^2(q,p),\\
    z^3(q+2\pi,p) = e^{i2\pi m^3/N}z^3(q,p)&,&
    z^3(q,p+2\pi) = e^{i2\pi n^3/N}z^3(q,p).
    \label{BDO}
   \eea
   Below, introducing $m^1 = - m^2 - m^3$, $n^1 = - n^2 - n^3$.
   solution to Eqs.~(\ref{BDO}) can be simply written as
   \be \label{BDO2}
    z^a(q,p) = e^{i(m^aq+n^ap)/N}f^a(q,p)
   \ee
   where $f^a(q,p)$ are periodic functions in $q$ and $p$.

   The membrane in Matrix Theory is ``quantized" by the
   prescription first to cut off all the components with
   the frequency higher than certain $K$ in the Fourier
   series of any world-volume function and then to substitute
   the two algebraic basis functions $e^{iq}$, $e^{ip}$ with
   the clock and shift matrices $U_K$ and $V_K$. Keeping
   track of the tensorial structure in the orbifolding,
   we are motivated to promote Eq. (\ref{BDO2}) to
   the Matrix Ansatz:
   \be\label{MMAP}
     Z^a(U_K,V_K) = U_K^{m^a/N} V_K^{n^a/N} F^a
     (U_K,V_K)\otimes \hat{V}_a
   \ee
   with $F^a$ polynomials in $U_K$ and $V_K$. \footnote{We are not
   bothered with commensurability of $N$ with respect to $m^a$ and
   $n^a$, because eventually $N$ is taken to infinity while keeping
   $m^a$ and $n^a$ finite. See ref. \cite{Bars} also.}

   To make sense of Eq. (\ref{MMAP}), we define the fractional
   powers of clock and shift matrices $U_K$, $V_K$ with ``nice''
   properties, a key technicality in this work. For mathematical
   rigor, we apply the {\em Dunford functional calculus}
   \cite{DUN} to matrices with finite rank $K$ by defining
   \bea
    \label{FP1}
      U_K^{a/c} &=& \frac{1}{2\pi i} \oint\limits_\Gamma
      \zeta^{a/c} (\zeta - U_K)^{-1} d\zeta,\\
      V_K^{b/d} &=& \frac{1}{2\pi i} \oint\limits_{\Gamma^\pr}
      \zeta ^{b/d} (\zeta - V_K)^{-1} d\zeta
    \label{FP2}
   \eea
   where $a, b, c$ and $d$ are arbitrary integers.
   From Eq.~(\ref{CS}), it is easy to show that the
   spectrum of $U_K$ contains all of the $K$-th roots
   of unity, $\omega_K^j$ for $j=0,1,\ldots, K-1$.
   To single out this spectrum, the contour $\Gamma$
   by definition consists $K$ disjoint small circles,
   each encircling an eigenvalue $\omega_K^j$, say
   $|\zeta - \omega_K^j|=\epsilon$ for some small
   $\epsilon>0$. A cut on $\zeta$-plane, running from
   the origin to the infinity, is drawn to sort out
   an analytic branch of the function $\zeta^{a/c}$.
   The cut can not have any intersections with the
   contour $\Gamma$, say passing between two neighboring
   circles. The same rules on the contour and cut should
   be applied for the definition of $\Gamma^\pr$ as well.
   Note that, as Eq.~(\ref{CS}), the definition in
   Eqs.~(\ref{FP1}) and (\ref{FP2}) is independent of
   specific matrix representation for $U_K$ and $V_K$.

   The following properties are generic results
   from the Dunford calculus \cite{Dunford}.
   \bea
    \label{ADD}
    U_K^{a/c} \cdot U_K^{a^\pr/c^\pr} = U_K^{a/c + a^\pr/c^\pr}&,&
    V_K^{b/d} \cdot V_K^{b^\pr/d^\pr} = V_K^{b/d + b^\pr/d^\pr},\\
    \label{HERM}
    U_K^{a/c\dag} = (U_K^\dag)^{a/c}&,&
    V_K^{b/d\dag} = (V_K^\dag)^{b/d},\\
    \label{COMP}
    (U_K^{a/c})^{a^\pr/c^\pr} = U_K^{aa^\pr/cc^\pr}&,&
    (V_K^{b/d})^{b^\pr/d^\pr} = V_K^{bb^\pr/dd^\pr}.
   \eea
   For example, Eq.~(\ref{ADD}) follows by manipulating Cauchy's
   integral formula and adopting the resolvent identity
   \be
    (\zeta^\pr-\zeta)^{-1}[(\zeta - U_K)^{-1}
    - (\zeta^\pr - U_K)^{-1}]
    =(\zeta - U_K)^{-1}(\zeta^\pr - U_K)^{-1}
   \ee
   (and a similar formula for $V_K$). A direct consequence of
   Eqs.~(\ref{HERM}) and (\ref{COMP}) is
   \be\label{UNITARY}
    U_K^{a/c\dag} = U_K^{-a/c},
    V_K^{b/d\dag} = V_K^{-b/d},
   \ee
   namely $U_K^{a/c}$ and $V_K^{b/d}$ are unitary.

   Because of the first two formulas in Eq.~(\ref{CS}),
   one has
   \be
      (\zeta - U_K)^{-1} = (\zeta^K - 1)^{-1}
     \sum\limits_{j=1}^K \zeta^{K-j} U_K^{j-1},\,
     (\xi - V_K)^{-1} = (\xi^K - 1)^{-1}
     \sum\limits_{j=1}^K \xi^{K-j} V_K^{j-1};
   \ee
   and therefore,
   \be
   \label{FP}
     U_K^{a/c} = \frac{1}{2\pi i} \sum\limits_{j=1}^K
       \oint\limits_\Gamma
       \frac{\zeta^{a/c+K-j}}{\zeta^K - 1}U_K^{j-1}d\zeta,\,
       \nonumber \\
     V_K^{b/d} = \frac{1}{2\pi i} \sum\limits_{j=1}^K
       \oint\limits_{\Gamma^\pr}
       \frac{\xi^{b/d+K-j}}{\xi^K - 1}V_K^{j-1}d\xi.
   \ee
   With the help of these equations, we have the following
   theorem:
   \begin{theorem}\label{TH4}
   \be
    \label{COMM}
      V_K^{b/d} \cdot U_K^{a/c}
     = \omega_K^{ab/cd} U_K^{a/c} \cdot V_K^{b/d}.
   \ee
   \end{theorem}
   The proof of this theorem is presented in Appendix~\ref{APP2}.

   This theorem is the central result of this subsection.
   It is a very nice property of the fractional powers we
   have defined. Comparing Eq. (\ref{COMM}) with the
   commutation relation $V_KU_K=\omega_KU_KV_K$, we see
   that the complex factor $\omega_K^{ab/cd}$ in the
   former is just a fractional power of $\omega_K$ in the
   latter. This property is highly non-trivial, because
   here we are dealing with the commutation relations of
   the fractional power of two {\em noncommuting} operators.
   \footnote{See \cite{MCSA} for the fractional powers of
   operators along a line other than the Dunford calculus.}

   After these preparations, now we can proceed with our
   Matrix Ansatz (\ref{MMAP}) for the wrapping states of
   a matrix membrane, which is a proper, noncommutative
   generalization of the ordinary wrapping map (\ref{BDO2}).
   Our discussion below on membrane physics will heavily
   rely on the commutation relation (\ref{COMM}).

   The next step to define those matrix states is to specify
   the matrix functions $F^a(U_K,V_K)$ in Eq.~(\ref{MMAP}).
   In this work, we will restrict ourselves to the
   center-of-mass motion, suppressing the oscillation modes.
   Therefore, $F^a$ are just complex numbers. As for the
   values for $m^a$, $n^a$, we take Schwarz' Ansatz
   \be
   \label{STT}
    n^2 = q_1, \qquad n^3 = q_2\,
   \ee
   whose reason will be explored in full length in Subsection~\ref{PREF}.
   Because of Eq.~(\ref{UNITARY}), motion in the radial
   directions and in the unorbifolded angular direction
   are suppressed, exactly the same as from Eq.~(\ref{METRIC})
   to Eq.~(\ref{METRIC1}). In accordance with subsection
   \ref{SUBSECTION1}, if we further require
   \be \label{FA}
     |F^a|=\bra z^a\ket
   \ee
   with $\bra z^a\ket$ are the VEV given in Eq.(\ref{VEV}),
   then the Matrix Ansatz (\ref{MMAP}) describes a matrix
   membrane wrapping on $T^2$, as dictated in Eq.~(\ref{STT}).
   In other words, contrary to Eq.~(\ref{DECOM}), the factor
   $F^a U_K^{m^a/N} V^{n^a/N}$ in Eq. (\ref{MMAP}) provides
   a polar decomposition of the bi-fundamental variable $Z^a$.
   Here $|F^a|$ may be interpreted as the distance from the
   membrane to the orbifold singularity (as center-of-mass
   degrees of freedom), while the unitary matrix
   $U_K^{m^a/N} V_K^{n^a/N}$ describes how the constituent
   D0-branes of the membrane are wrapped in the orbifolded
   angular direction of $Z^a$.

   \subsection{Dynamics of Wrapping Membranes}

   After defining a class of wrapping membrane states
   with Eqs.~(\ref{MMAP}), (\ref{STT}) and (\ref{FA}),
   we discuss their physics in this subsection. Since
   the wrapping states involves pulling the geometry of
   $T^2$ back to the membrane in subsection \ref{WRAPPING},
   the membrane probes the geometry constructed in
   subsection.~\ref{SUBSECTION1} via the wrapping states.

   \subsubsection{Classical Motions}

   As part of the BFSS conjecture \cite{BFSS}, membranes
   in Matrix Theory are considered to be a composite of
   D0 branes. Therefore, the action (\ref{OA}) of our
   quiver matrix mechanics, as an orbifolded Matrix Theory,
   legitimately describes the dynamics of the matrix
   membrane degrees of freedom for the states given by
   Eq.~(\ref{MMAP}). Generically the center-of-mass
   degrees of freedom $F^a$ are time dependent. The
   classical motion for $F^a(t)$ in Eq.~(\ref{MMAP})
   is determined by the equation of motion (EOM)
   derived from the action (\ref{OA}):
   \be\label{EN}
   \ddot{Z^a} + \frac{R_{11}^2}{2}([Z^b,[Z^{b\dag}, Z^a]]
   +[Z^{b\dag},[Z^b,Z^a]]) = 0,
   \ee
   in which we have recovered $R_{11}$ explicitly.
   For convenience, we write $Z^a(U_K,V_K)$ shortly
   as $Z^a$ hereafter without confusion.
   Eq.~(\ref{EN}) is satisfied if
   \be
   \label{SOL}
    F^a(t)=\bra z^a\ket e^{-i\omega_a t}
   \ee
   with
   \be
   \label{OMEG}
   \omega_a^2 = ({R_{11} N\over 2\pi})^2
   (1-\cos {2\pi w\over KN^2})
   \sum\limits_{b\neq a} |f_b|^2.
   \ee
   Moreover, the solution (\ref{SOL}) also solves an additional
   constraint
   \be\label{CONS}
    [\dot{Z}^a,Z^{a\dag}]+[\dot{Z}^{a\dag},Z^a]=0,
   \ee
   that descends from the gauge fixing of the membrane world-volume
   diffeomorphism (see for example \cite{Taylor}).

  \subsubsection{Wrapping Spectrum}

   First let us calculate the energy density on a wrapping membrane:
   \be
   \label{ENGD}
    \H = {R_{11}\over 2}
    \{|[Z^a,Z^{b\dag}]|^2 + |[Z^a,Z^b]|^2\}.
   \ee
   Taking into account Eqs.~(\ref{MMAP}) and (\ref{SOL}), one has
   \be
     \H = R_{11}({N\over 2\pi})^4 (f_1^2f_2^2+f_1^2f_3^2+f_2^2f_3^2)
     (1-\cos {2\pi w\over KN^2}) \unit_{KN^2}.
   \ee
   Instead of Eq.~(\ref{TRACE}) in the context of the SYM
   limit, for wrapping membrane states the trace of
   $\unit_{KN^2}$ is regularized to be $(2\pi)^2K$.
   Recalling the toroidal metric (\ref{METRIC2}),
   in either the large $N$ or the large $K$ limit,
   the wrapping energy approaches to
   \be\label{WE}
    P_w = Tr\H = \frac{M_w^2}{2P^+}
   \ee
   where
   \be\label{MW}
    M_w = T_{M2} w A_{T^2}
   \ee
   with $T_{M2}=1/(2\pi)^2$ the dimensionless membrane
   tension, $A_{T^2}$ the area of $T^2$ given by
   Eq.~(\ref{AREA}), and the light-cone momentum
   \be\label{PP}
    P^+ = K/R_{11}.
   \ee
   Eqs.~(\ref{WE}), (\ref{MW}) and (\ref{PP}) match
   perfectly with the M-theory picture. The light-cone
   energy $P_w$ is of the nonrelativistic form in
   Eq.~(\ref{WE}), with light-cone mass $P^+$; the
   transverse (wrapping) mass $M_w$ is factorized
   exactly into the correct membrane tension,
   wrapping number and the area of the torus in target
   space. As a finite energy state, the light-cone
   energy scales like $\order(1/K)$, as predicted by
   BFSS \cite{BFSS}.

   \subsubsection{Stability of Configuration}

   From Eq.~(\ref{OMEG}), the configuration in Eq.~(\ref{SOL})
   is static if and only if there is no wrapping (provided
   $w\ll N$.) This is not a surprise that a wrapped static
   membrane cannot stay stably on an orbifold because,
   the closer the membrane to the orbifold singularity,
   the less the tension energy costs in Eq.~(\ref{MW}). (In
   other words, the only stable static wrapping configuration
   is the one in which all D-particles stay right at the origin.)

   Consequently, a wrapping membrane away from the origin has
   to rotate to achieve a stationary state. The rotation with
   the angular velocity $\omega_a$ contributes a nonzero
   kinetic energy:
   \be\label{ENG}
    K_w = \frac{1}{R_{11}}Tr|\dot{Z}^a|^2
    = P_w.
   \ee

   Another observation is
   \be
   |\omega_a| = R_{11} {N\over \pi} \sin({\pi w\over KN^2})
   [\sum\limits_{b\neq a} f_b^2/2]^{1/2}.
   \label{OMEGS}
   \ee
   The interpretation of Eq.~(\ref{OMEGS}) is simple:
   due to the fuzziness introduced by finite $N$, the
   wrapping membrane generally rotates also in the
   un-orbifolded direction $\varphi^1$.

   \subsubsection{Center-of-Mass Momenta}

   Eq.~(\ref{MMAP}) is not the most general solution to
   the equations of motion (\ref{EN}) and (\ref{CONS}).
   At least one can add a term linear in time:
   \be
   \label{MM}
Z^a = {f_a\over \sqrt{2} 2\pi}(N e^{-i\omega_a t}
    U_K^{m^a/N} V_K^{n^a/N}
    + {iR_{11}\over K} k^a t ) \otimes \hat{V}_a \, ,
   \ee
   where real coefficients $k^a$ are to be determined.
   Now the total (light-cone) energy of the configuration
   (\ref{MM})at finite $N$ and $K$ is
   \be\label{ENER}
    H_{mem} = \frac{1}{2P^+} {tr\over K}
    \{f_a^2 |k^a - {\sin \pi w/KN^2 \over \pi/KN^2}
[\sum\limits_{b\neq a} f_b^2/2]^{1/2}
    e^{-i\omega_a t}U_K^{m^a/N} V_K^{n^a/N}|^2\} + P_w.
   \ee
   To evaluate Eq.~(\ref{ENER}), we need to know
   $tr \{U^{m^a/N} V^{n^a/N}\}/K$. From Eq.~(\ref{FP})
   and the definition of the contours $\Gamma$,
   $\Gamma^\pr$ and the branch cuts, we get
   \bea\nonumber
    {tr\over K} \{U^{m^a/N} V^{n^a/N}\} &=&
{1\over K^2}(\sum\limits_{j=1}^K e^{i2\pi jm^a/KN})
    (\sum\limits_{j^\pr=1}^K e^{i2\pi j^\pr n^a/KN})\\
&=& \frac{e^{i2\pi m^a/KN}
    (e^{i2\pi m^a/N} - 1)}{K(e^{i2\pi m^a/KN} - 1)}\cdot
\frac{e^{i2\pi n^a/KN}(e^{i2\pi n^a/N} - 1)}
    {K(e^{i2\pi n^a/KN} - 1)}.
   \eea
   So in the large-$K$ and large-$N$ limit (continuous
   membrane and continuous torus limits), with $R_{11}=K$,
   \be
    H_{mem} = {1\over 2} (f_a^2p^{a2} + M_w^2)
   \ee
   where $p^a = k^a - w[\sum\limits_{b\neq a} f_b^2/2]^{1/2}$.
   From the point of view of $T^2$,
   we require $p^1 = - p^2 - p^3$;
   therefore $k^a$ can not be independent.
   And we also require that the canonical momenta
   $p_\alpha = g_{\alpha\beta} p^\beta$ are quantized to
   take integer values $l_2$, $l_3$; accordingly,
   \be\label{ME}
    H_{mem} = {1\over 2} (g^{\alpha\beta}l_\alpha l_\beta + M_w^2).
   \ee

   \subsection{IIB/M(atrix) Duality from Wrapping Membranes}

   Now we are in a position to verify the IIB/M(atrix) duality
   by studying the spectrum and symmetry of matrix membrane
   states.

   \subsubsection{Elimination of Unwanted Degeneracy}
   \label{PREF}

   Generically different wrapping map matrices may have the
   same wrapping number. Since the energy of wrapping states
   is proportional to $w$, we seem to encounter a possible
   enormous degeneracy for a given $w$. Even the $SL(2,\integer)$
   equivalence in Eq.~(\ref{SL2Z}) is not able to resolve all
   the degeneracy. On the other hand, if one wants to make
   the correspondence between wrapping membrane states and
   the doubly-charged $(q_1,q_2)$ string states, in accordance
   with IIB string theory, the wrapping membranes with a given
   wrapping number $w$ should be non-degenerate \cite{SCH2}.
   To eliminate the degeneracy of wrapping membranes, Schwarz
   incorporates the Kaluza-Klein direction in to picture
   for the $(q_1,q_2)$ strings \cite{SCH2}.

   According to the generalized T-duality that we briefed
   in Sec.~\ref{section_int}, $(q_1,q_2)$-string winding
   $l$ times on the IIB theory circle is dual to KK mode
   $(lq_1,lq_2)$ on $T^2$ in M-theory, with $q_1$, $q_2$
   coprime. Schwarz' {\it ansatz} says that one cycle of
   the membrane, say, in $p$-direction, {\em must} posit
   in the KK direction and winds only once, namely
   $\vec{n} = (q_1,q_2)^T=:\vec{q}$. Accordingly,
   \be\label{WN}
    w = m^2q_2-m^3q_1 \, .
   \ee
   We claim the following theorems:
   \begin{theorem}
   \label{TH1}
   For any pair of coprime integers $(q_1,q_2)$,
   there {\em exists} a pair of integers $(m^2,m^3)$
   such that Eq.(\ref{WN}) is satisfied with $w=1$.
   \end{theorem}
   This theorem is an elementary result in Number Theory;
   it has an (obviously) equivalent presentation:
   \begin{theorem}
   \label{TH3}
   For any integer $w$ and any pair of coprime integers
   $(q_1,q_2)$, there exists a pair of integers $(m^2,m^3)$
   such that Eq.(\ref{WN}) is satisfied.
   \end{theorem}
   The following theorem asserts the uniqueness of
   the considerations:
   \begin{theorem}\label{TH2}
   If there exists another pair $(m^{\pr 2},m^{\pr 3})$
   satisfying Eq.~(\ref{WN}), then there {\em exists}
   an $SL(2,\integer)$ transformation on the membrane
   coordinates $(q,p)$ that relates these two wrapping
   maps.
   \end{theorem}
   We present a proof to Theorem~\ref{TH2} here.
   Because both $(m^2,m^3)$ and $(m^{\pr 2},m^{\pr 3})$
   satisfy Eq.~(\ref{WN}),
   $(m^{\pr 2} - m^2)q_2 = (m^{\pr 3} - m^3)q_1$.
   Subsequently,
   there exists an integer $b$ such that
   $m^{\pr 2} = m^2 + bq_1$,
   $m^{\pr 3} = m^3 + bq_2$.
   Then,
   as usual $\vec{m^\pr}:=(m^{\pr 2},m^{\pr 3})^T$,
   \be
    W(\vec{m^\pr},\vec{q}) =
    W(\vec{m},\vec{q})
    \left(
     \begin{array}{cc}
      1 & 0 \\
      b & 1
     \end{array}
    \right).
   \ee
   {\it Q.E.D.}

   Using these theorems, the original characterization of a
   wrapping map is traded into three parts, $(q_1,q_2)$-charge,
   wrapping number $w$ and an $SL(2,\integer)$ family labelled by $b$. Thus the
   wrapping states with given $w$ is non-degenerate up to the
   geometric $SL(2,\integer)$ symmetry on the membrane.

  \subsubsection{IIB/M Duality}

   To see how the generalized T-duality \cite{SCH2} works,
   on the one hand $w$ is the wrapping number of a membrane
   over $T^2$ in M-theory; on the other hand the wrapping
   mass (\ref{MW}) can be reinterpreted as a KK momentum
   as suggested by IIB/M duality, namely
   \be
    M_w = w/R_B,
    R_B = 1/\sqrt{g}
   \ee
   where the newly constructed IIB circle $S_B$ has radius
   $R_B$. $S_B$ becomes decompactified when the size of
   $T^2$, measured with $\sqrt{g}$, shrinks to zero;
   hence, the wrapping contribution to the spectrum (\ref{ME})
   becomes a continuous kinetic energy. To complete the
   generalized T-duality, besides Eq.~(\ref{STT}), we add the
   following known constraint on the center-of-mass momenta
   of the membrane, coming from the argument of the stability
   of the FD-string bound states, that
   \be
    l_2 = l q_1,
    l_3 = l q_2.
   \ee

   The counterpart of the ``nine-eleven flip" in IIA/M duality
   is the identification of the modular parameter $\tau$ of
   $T^2$ and the IIB coupling $\chi + i e^{-\phi}$, where $\chi$
   is the Ramond-Ramond scalar and $\phi$ the dilaton in IIB
   string theory. In accordance with this identification,
   the metric of $T^2$ can be expressed in IIB terms as
   \be\label{MTRC}
    (g_{\alpha\beta}) = e^{2\omega}
    \left(
     \begin{array}{cc}
      1&\chi \\ \chi&\chi^2+e^{-2\phi}
     \end{array}
    \right).
   \ee
   Therefore, the parametrization $f_a$ is just the moduli of
   an RR-scalar, dilaton and an overall radion \cite{Kaplan}.

   Since both the determinant $g$ and the wrapping number
   are invariant under $SL(2,\integer)$ transformation of
   $T^2$, so is the wrapping mass (\ref{MW}). Moreover,
   $(q_1,q_2)$ transform covariantly under $SL(2,\integer)$,
   so the kinetic energy (\ref{ME}) is also invariant,
   as well as the total energy $H_{mem}$.

  \section{Conclusions and Discussions}
  \label{section_dis}

   The logic underlying this paper can be summarized as follows.
   First the geometry setting is the orbifold
   $\complex^3/\integer_N^2$, in which a discretized torus is
   ``embedded'' even at finite $N$. On the one hand, the collective
   motion of D-particles in the angular directions on this
   orbifold may develop a (regularized) wrapping membrane,
   which are described by fractional powers of the clock and
   shift matrices. Our quiver matrix mechanics governs the
   dynamics of the wrapping configurations. The wrapping
   modes develop a Kaluza-Klein tower, giving rise to the
   generalized T-duality and IIB/M(atrix) duality.
   On the other hand, by deconstruction of D-particle states,
   the dual torus emerges in target space and $1+2$-dimensional
   SYM emerges in the large $N$ limit. This deconstruction
   procedure also reveals a (hidden) underlying $1+3$-dimensional
   SYM, which plays a vital role in the literature of
   IIB/M(atrix) Theory duality.

   Our analysis of IIB/M duality concentrates mainly on the
   spectroscopy, leaving for the future research of the deduction
   of an effective theory from SYM as well as the relation
   between Yang-Mills coupling and IIB coupling. A spectroscopic
   discussion of wrapping membrane is also given in \cite{KO},
   where $T^3\times A_{N-1}$ is taken to be the base space.
   We only note that to get the correct chirality of IIB
   fermions from the Matrix Theory is highly nontrivial.
   The deconstruction technique has been widely employed in
   string community, for examples see \cite{RS,ACKKM};
   more comprehensive discussion on compactification in
   various dimensions, especially on M5-brane, can be found
   in \cite{FHRS,OGSS} (see also \cite{Ganor} and \cite{KO_e}
   for intersectional M5-brane). As we noted before, the main
   technical difficulty that we have overcome is to construct
   the matrix membrane states wrapping on the orbifold. Our
   formalism of fractional powers of clock and shift matrices
   has a natural connection with fractional membrane, of
   which early intuition can be traced to \cite{DDG}.

   We have applied our approach to both IIA/M and IIB/M
   dualities. While IIA/M duality is now a somewhat
   common exercise\footnote{For the latest discussions,
   see \cite{HI,UY}.}, the success of our quiver matrix
   mechanics approach in demonstrating IIB/M duality
   shows its powerfulness in dealing with non-perturbative
   aspects of string theory, in contrast to the inaptitute
   of compactified Matrix Theory to incorporate the wrapping
   matrix membrane states.


  {\bf Acknowledgement}

  JD thanks the High Energy Astrophysics Institute and Department
  of Physics, the University of Utah, and Profs. K. Becker, C. DeTar
  and D. Kieda for warm hospitality and financial support; he also
  appreciates Li-Sheng Tseng for helpful discussion on some
  mathematical aspect of this work. YSW thanks the Interdisplinary
  Center for Theoretical Sciences, Chinese Academy of Sciences,
  Beijing, China for warm hospitality during his visit, when the
  work was at the beginning stage.

  \appendix
  \section{Proof of Theorem~\ref{TH4}}
  \label{APP2}

   We have to show the following lemma first.
   \begin{lemma}\label{LEM}
   For arbitrary integer $j$,
   \be\label{LEMMA}
    (\omega_K^j U_K)^{a/c} = \omega_K^{ja/c} U_K^{a/c};
   \ee
   a similar statement holds for $V_K^{b/d}$ too.
   \end{lemma}
   In fact,
   \bea
    \nonumber
\mbox{R.H.S. of Eq.~(\ref{LEMMA})} &=& \frac{1}{2\pi i}
      \sum\limits_{j^\pr = 1}^K \oint\limits_\Gamma
\frac{(\omega_K^j\zeta)^{a/c}\zeta^{K-j^\pr}}
  {\zeta^K - 1}U_K^{j^\pr - 1}d\zeta\\
&=& \frac{1}{2\pi i} \sum\limits_{j^\pr = 1}^K \oint\limits_\Gamma \frac{(\omega_K^j\zeta)^{a/c+K-j^\pr}}
    {(\omega_K^j\zeta)^K - 1}
(\omega_K^jU_K)^{j^\pr - 1}d(\omega_K^j\zeta).
   \eea
   In the last line, we used the fact that
   $(\omega_K^j)^K = 1$. Changing the variable from
   $\zeta$ to $\omega_K^i\zeta$ just shifts cyclically
   the circles constituting the symmetric contour
   $\Gamma$, without the integral unchanged.
   Recall the definition in Eq.~(\ref{FP}), and
   the L.H.S. of Eq.~(\ref{LEMMA}) follows.

   Then the proof of Eq.~(\ref{COMM}) becomes straightforward.
   By substituting Eq.~(\ref{FP}),
   \be\label{LHS}
    \mbox{L.H.S. of Eq.~(\ref{COMM})}=
    \frac{1}{2\pi i} \sum\limits_{j=1}^K \oint\limits_{\Gamma^\pr}
    \frac{\xi^{b/d+K-j}}{\xi^K - 1}
    \frac{1}{2\pi i} \sum\limits_{j^\pr=1}^K \oint\limits_\Gamma
    \frac{\zeta^{a/c+K-j^\pr}}{\zeta^K - 1}
    V_K^{j-1}U_K^{j^\pr-1}
    d\zeta d\xi.
   \ee
   Because of the commutation relation in Eq.~(\ref{CS}),
   $V_K^{j-1}U_K^{j^\pr-1} =
   \omega_K^{(j-1)(j^\pr - 1)}U_K^{j^\pr-1}V_K^{j-1}$.
   Accordingly, the R.H.S of Eq.~(\ref{LHS}) is
   \be\label{LHS1}
    \frac{1}{2\pi i} \sum\limits_{j=1}^K \oint\limits_{\Gamma^\pr}
    \frac{\xi^{b/d+K-j}}{\xi^K - 1}
    [
    \frac{1}{2\pi i} \sum\limits_{j^\pr=1}^K \oint\limits_\Gamma
    \frac{\zeta^{a/c+K-j^\pr}}{\zeta^K - 1}
    (\omega_K^{j-1}U_K)^{j^\pr-1}
    d\zeta ] V_K^{j-1} d\xi.
   \ee
   By definition (\ref{FP}), $[\ldots ]$ in Eq.~(\ref{LHS1}) is
   just $(\omega_K^{j-1}U_K)^{a/c}$. Due to Lemma~\ref{LEM},
   $(\omega_K^{j-1}U_K)^{a/c} = \omega_K^{(j-1)a/c}U_K^{a/c}$.
   Then, the R.H.S of Eq.~(\ref{LHS1}) is
   \be
   \label{LHS2}
U_K^{a/c} \frac{1}{2\pi i} \sum\limits_{j=1}^K \oint\limits_{\Gamma^\pr} \frac{\xi^{b/d+K-j}}{\xi^K - 1}
    (\omega_K^{a/c}V_K)^{j-1} d\xi.
   \ee
   Again by definition in Eq.~(\ref{FP}),
   Eq.~(\ref{LHS2}) gives rise to
   $U_K^{a/c}(\omega_K^{a/c}V_K)^{b/d}$.
   Finally, using Lemma~\ref{LEM} again,
   the R.H.S. of Eq.~(\ref{COMM}) follows. {\it Q.E.D.}


 \end{document}